# White-Light Imaging in a Two Gratings Diffraction Process

**José J. Lunazzi and Noemí I. R. Rivera**\*

*Universidade Estadual de Campinas, Instituto de Física Gleb Wataghin, Caixa Postal 6165,*

*13083-970 Campinas, SP, Brazil*

## Abstract

A diffractive arrangement that allows imaging of an object without any intermediate or complementary element is presented. This optical system with only two diffraction gratings forms color images with white light.

*Keywords*: Diffractive Optics; Three-dimensional image; Image System; Holography.

\* The authors can be reached by e-mail as follows: José J. Lunazzi, lunazzi@ifi.unicamp.br, and Noemí I. Rivera, nrivera@ifi.unicamp.br.

---

[1] This paper was rejected by Journal of Pure and Applied Optics, even after many appeals, as being of no value: "it presents low quality work that adds little or no new knowledge to the field." 2005-2006
[2] This paper was also rejected by Applied Optics' editor James Wyant as being merely the Talbot effect. 2006
[3] This paper was considered not appropiated for Journal of Modern Optics due to lacking of application of the paper in some technical matter. 2006
[4] This paper was not considered appropiated for publication in Nature Photonics. 2007
[5] Rejected for publishing at RIAO/OPTILAS 2007, October 21-26, 2007 Campinas Sao Paulo Brazil

# 1. INTRODUCTION

There are reports of interesting hybrid system developments where the double diffraction imaging is intermediated by a refractive element [1]. Imaging with only diffracting elements however, cannot be found in the literature, and a paper by Bennet [2] indicates the impossibility of achieving white light imaging with just two holographic elements. In the space telescope project named "Eyeglass" [3], two spaceships are proposed, one carrying a large thin diffractive telescope objective and the other containing a refractive objective and corrective diffracting elements. The first experimental results on this technique were presented by Dixit [4] who indicated that the elimination of the refractive element in the telescopic imaging process would be a further evolution because glass objectives are heavy loads to spaceships [5]. The present work describes a double-diffractive arrangement that entails the elimination of the refractive element. Bennet [2] demonstrates the impossibility of white-light images when using paraxial approximations and holographic elements. We circunvent this impossibility by working at large angles. Our arrangement, however, does not perform a convergent image.

Previous works have demonstrated that a double diffraction system intermediated by a slit yields images that were perfectly symmetric and consequently depth inverted (pseudoscopics) [6-8]. This same system shows images with normal depth (orthoscopic images) when the observer looks at a symmetrical diffraction order of the second diffraction process [9,10]. Moreover, properties of images for white-light objects using two bi-dimensionally diffractive elements and an intermediary pinhole has been demonstrated [11,12]. All these previous results were derived from symmetric situations, for which the enlarging of the aperture of the intermediate element extends the image point to an aberration spot. It is therefore reasonable to assume that the intermediate element has a fundamental role. Nevertheless, we found [13] that eliminating this element yields an interesting image that results when the first diffractive element has half the period of the second – i.e., the spreading of emerging light from the first diffraction can be compensated by the second diffraction process. Preliminary results employing two bi-dimensionally defined diffracting elements (spiral gratings) show the possibility of obtaining a divergent image [14].

It might be opportune to explain why the results described in the present work cannot be associated with either the Talbot or the Lau effects [15-16] . In these effects, an infinite periodical element (a linear grating) illuminated under white light generates a converging self-image at certain distances from a second linear grating which acts like an imaging forming element. Thus, for the Talbot and Lau effects the object is necessarily one of the gratings while in the present results – and in those previuosly reported (5 to 13) – the object is independent of the imaging system and it is non-periodic. Besides, the Talbot and Lau effects need the combination of all diffraction orders while for both the present and previously reported works (5 to 13) one particular order from each diffracting element is selected at a time.

## 2. DESCRIPTION

The Figure 1 shows the configuration used to observe white-light images with a double diffraction process without an intermediate refractive element. In this three-dimensional figure, the white-light point object $O$ is at distance $Z$ from the first diffractive grating diffractive element DG1 with $2\nu$ lines per millimeter and normal to the figure plane. The luminous rays exiting $O$ impinges DG1 at point $X_{1n}$, and the corresponding diffracted rays reach the second diffractive grating DG2 at the point $X_{2m}$ located at distance $Z_R$ from DG1. DG2 has $\nu$ lines per millimeter and its lines are also normal to the figure plane. The resulting double-diffracted rays convey the characteristics presented below. The image is seen either by an observer or by a camera having the pupil diameter approximately equal to the eye. As shown in Fig. 1, the imaginary virtual rays that are the prolongation of the double-diffracted rays cross at the image $i$, $i_b$ being the image for a short wavelength and $i_r$ for a large wavelength. As presented below, a simple diffraction grating equation can be used to analyze the recombination capacity of the diffracted rays at the second diffraction element, i.e., the observation of an image that is located next to the object (see Fig. 1).

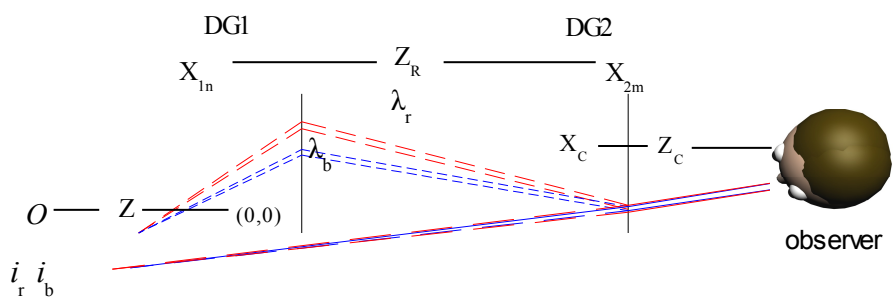

Figure 1. Imaging scheme of a double diffraction process without intermediated element. The virtual prolongation of the double-diffracted rays cross at the image $i$, $i_b$ being the image for a short wavelength and $i_r$ for a large wavelength.

The first-order diffraction at the first grating follows the equation:

$$\sin\theta_i - \sin\theta_d = 2\lambda \cdot \nu \qquad (1)$$

where $\theta_i$ is the angle of incidence, $\theta_d$ is the angle of diffraction, $\nu$ is the spatial frequency and $\lambda$ is any wavelength within the visible spectrum. Eq.1 can be expressed in terms of coordinates as:

$$-\frac{X_{1n}}{\sqrt{X_{1n}^2 + Z^2}} - \frac{X_{1n} - X_{2m}}{\sqrt{(X_{1n} - X_{2m})^2 + Z_R^2}} = -2\lambda_{nm}\nu \qquad (2)$$

where $\lambda_{nm}$ is the wavelength of each ray between $X_{1n}$ and $X_{2m}$.

Analogously, the first order diffraction at the second grating can be expressed as:

$$\frac{(X_{1n} - X_{2m})}{\sqrt{(X_{1n} - X_{2m})^2 + Z_R^2}} + \frac{(X_C - X_{2m})}{\sqrt{(X_C - X_{2m})^2 + Z_C^2}} = +\lambda_{nm}\nu \qquad (3)$$

where $X_C$, $Z_C$ are the observer's coordinates. The exit angle for the double diffracted rays is:

$$\Phi_{dn} = \arcsin\left[\frac{X_{1n}}{\sqrt{X_{1n}^2 + Z^2}} - \lambda_{nm}\nu\right]. \qquad (4)$$

The values $X_{1n}$ and $X_{2m}$ can be obtained from equations (2) - (3). Using these values in equation (5), one obtains the image position (Xi, Zi) for each wavelength $\lambda_b$ and $\lambda_r$. In the present analysis, $\lambda_b < \lambda_r$, thus the point images $i_b$ and $i_r$ in the Fig. 1 correspond to coordinates (Xi$_b$, Zi$_b$) and (Xi$_r$, Zi$_r$), respectively.

$$\frac{X_{1n}-X_{2m}}{\sqrt{(X_{1n}-X_{2m})^2+Z_R^2}} + \frac{X_{2m}-Xi_{b,r}}{\sqrt{(X_{2m}-Xi_{b,r})^2+Zi_{b,r}^2}} = \lambda_{nm} \nu \qquad (5)$$

The image keeps parallax in the two possible planes of Fig. 1, normal and parallel to the page. In the first diffraction plane, the parallel is only limited by the grating size, while in the second possible plane there exist a blurring effect due to the sequence of different images: each wavelength giving an image at a different lateral position.

## 3. EXPERIMENTAL DETAILS

The experimental set up consisted of two holographic transmission gratings of the same type, with 1080 ± 9 lines/mm for the first grating, and 503 ± 8 lines/mm for the second one. The effective area of the first grating was approximately 78 mm x 57 mm, and that of second 78 mm x 17 mm. These parallel gratings were located 90 ± 5 mm apart. As mentioned in the previous section of analysis, first-order diffraction rays of both gratings were employed, corresponding to the above set of Eqs. 1 to 3.

The two objects employed were 40 W filament lamp with its 0.5 mm wide filament oriented normal to the plane of Fig.1, and an extended object composed of a metal stamp with 36 mm x 36 mm dimensions whose surface was made to diffuse light with aluminum paint. This stamp was illuminated with a 50 W halogen lamp. Both objects were located 425 ± 5 mm from the first grating. Either the observer or the camera were located at a distance of 468 ± 5 mm to the second grating and its visual line made an angle of 6.8 degrees with the normal to the gratings. The aperture of the lens was 3 mm to keep it comparable to the aperture of an observer's eye.

## 4. RESULTS AND DISCUSSIONS

## 4.1 Selecting rays with a slit

The purpose of the first experimental tests described in this section was that of a selective analysis of the image process. This analysis was obtained by comparing the image of the filament lamp with that obtained with a selected wavelength bandwidth. This bandwidth was set with a 1 mm wide slit located at the first grating parallel to its lines. Figure 2 shows the photograph of the white-light image at left, and at its right-hand side is the image obtained in other available spectral bandwidths of the source.

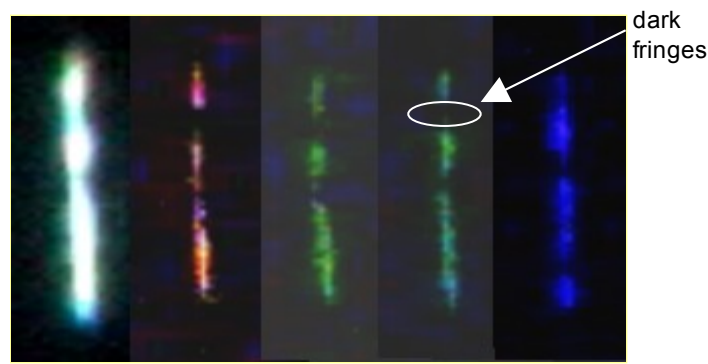

Figure 2. Images of the filament of a lamp. The leftmost is a direct white image. The other images were obtained with slit filtering.

In Fig. 2, the last three images to the right correspond to shorter wavelengths: the two in the middle in the green color range, and the last in blue. It can also be seen in Fig. 2 the distinct dark fringes cutting all the filtered images. These fringes are generated by the opaque characters on the lamp bulb surface, 27 mm in front of the filament. Thus, for a particular point-object source, the irradiating rays within having the wavelength window have different paths and intersect the opaque characters at different places. The dark horizontal fringes at different positions imply different perspectives of the same scene. This situation is comparable to that of a previously reported case of imaging through a diffraction grating [17], where each wavelength also yields an image with a different perspective.

## 4.2 Out of focus imaging

When considering diffraction at a grating (see Fig. 3) it should be noted that for each wavelength the observer sees a different perspective of the object because the collected rays leave the object at different angles. For a point or line object, the defocusing width on the image is proportional to the distance from the focused position – analogous to the case of refractive lens imaging. As shown in Fig. 3, in the present configuration the spectral sequence of colors is distinguishable, making it easier to identify or measure defocusing. This can be accomplished by illuminating the object with two or more well defined wavelengths – laser lines, for example.

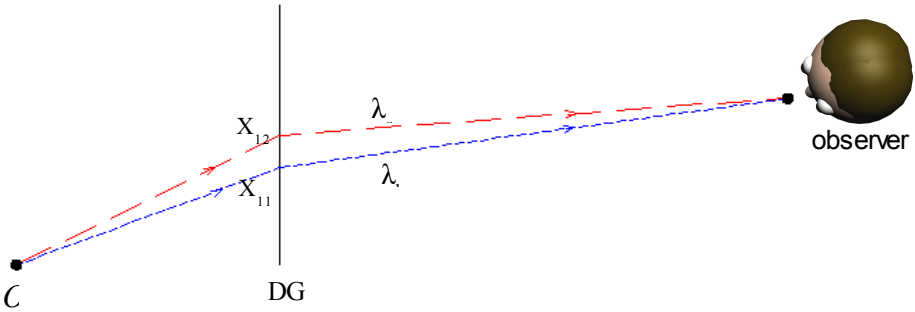

Figure 3. Scheme showing perspective variation associated with the wavelength value.

Figure 4 shows the defocusing when the object is focused 38 mm and 48 mm ahead or behind of the focused position, respectively. In a black and white reproduction the spectral colors cannot be noticed.

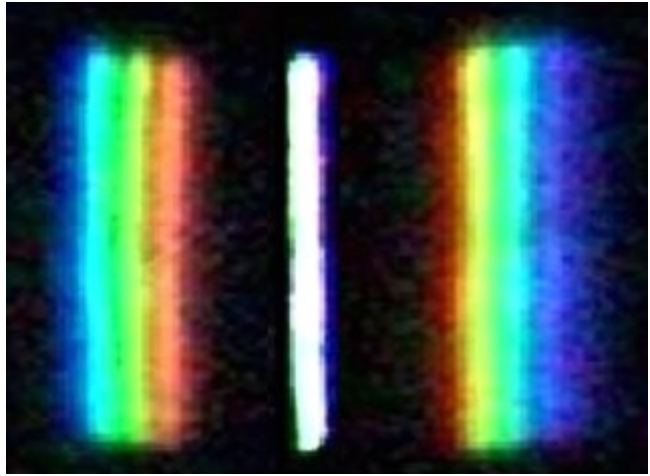

Figure 4. Defocusing effects on the object image. The three positions are: back-focused position (left image); focused position (center image); and front focused position (right image).

### 4.3 Image of an extended object

The present configuration also exhibits well defined view of an extended object. Figure 5 shows a comparison of images obtained with photography and the present arrangement, respectively. The object employed was the metal stamp described in section 3. Fig. 5 (left) shows the direct photograph of the object, while Fig. 5 (right) is the image obtained after double diffraction.

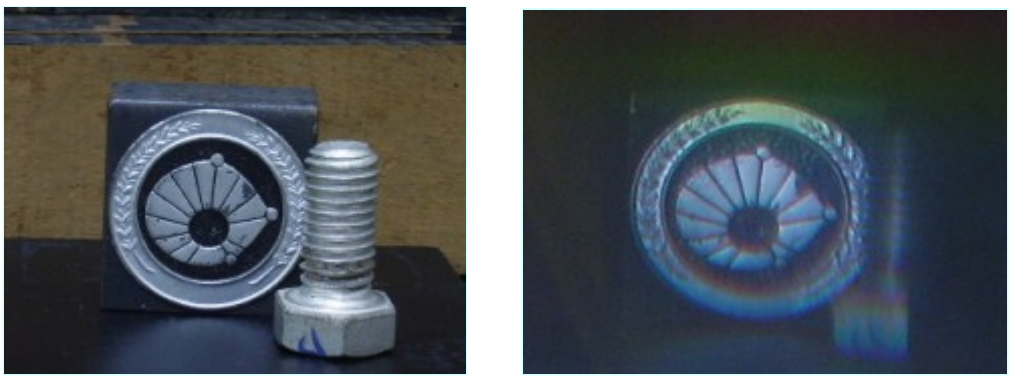

Figure 5. The image of an extended object. For comparison, the photograph of the extended object is shown at the left side. On the right side is the image of an extended object.

## 5. CONCLUSIONS

It was possible to obtain an image with an optical process that involves only double diffraction without any intermediate element. This new optical system renders images comparable to those of plane refractive elements but shows focusing properties. Since the second diffractive element has almost total transparency – it is not perceived by the observer. – the image gives the illusion of being a ghost image, therefore, of great attractive appearance. Besides, it encodes depth in a way that could be useful for remote metrology.

## ACKNOWLEDGEMENTS

The "Pro-Reitoria de Pós Graduação" of Campinas State University - UNICAMP is acknowledged for a BIG fellowship for Noemí I. Rodríguez Rivera.